\documentclass[lettersize,journal]{IEEEtran}

\usepackage{graphicx} 
\usepackage{xcolor} 
\usepackage{cite}
\usepackage{amsmath,amssymb,amsfonts}
\usepackage{authblk}
\usepackage{float}
\usepackage{cuted}
\usepackage{mathtools}

\newcommand{\dom}{\mathrm{d}\omega}

\title{Shannon Theory for Wireless Communication in a Resonant Chamber}
\author{\small Amritpal Singh --- as14141@nyu.edu, Thomas L. Marzetta --- tom.marzetta@nyu.edu 
\thanks{Research supported by NYU WIRELESS.

This work has been submitted to the IEEE for possible publication. Copyright may be transferred without notice, after which this version may no longer be accessible.
}}
\date{March 2023}

\begin{document}

\maketitle
\begin{abstract}
    A closed electromagnetic resonant chamber (RC) is a highly favorable artificial environment for wireless communication. A pair of antennas within the chamber constitutes a two-port network described by an impedance matrix. We analyze communication between the two antennas when the RC has perfectly conducting walls and the impedance matrix is imaginary-valued. The transmit antenna is driven by a current source, and the receive antenna is connected to a load resistor whose voltage is measured by an infinite-impedance amplifier. There are a countably infinite number of poles in the channel, associated with resonance in the RC, which migrate towards the real frequency axis as the load resistance increases. There are two sources of receiver noise: the Johnson noise of the load resistor, and the internal amplifier noise. An application of Shannon theory yields the capacity of the link, subject to bandwidth and power constraints on the transmit current. For a constant transmit power, capacity increases without bound as the load resistance increases. Surprisingly, the capacity-attaining allocation of transmit power versus frequency avoids placing power close to the resonant frequencies.
\end{abstract}

\section{Introduction}
\IEEEPARstart{C}{ommunication} systems are fundamentally limited by the physics of wave propagation. Therefore, communication theorists can benefit from placing greater emphasis on natural phenomena. Analyzing the number of degrees of freedom of a radiating field as well as determining channel capacity are two avenues among others to study information theory while acknowledging the underlying physics of radiation \cite{franceschetti2017wave,  migliore2008electromagnetics,migliore2018horse}. Here, we take the latter approach to electromagnetic information theory by obtaining the Shannon capacity of a resonant chamber.

The resonant chamber constitutes a highly favorable artificial propagation environment that simultaneously provides isolation from the exterior environment (no interference to the exterior and no interference from the exterior), and propagation that exhibits ``rich scattering." Potential applications include robot-equipped factories, and large data centers in which servers are connected wirelessly.

The resonant chamber (RC) --- or resonant ``cavity" --- is an enclosed space with a highly conductive boundary. To be clear, this work is not referring to the closely related ``reverberation" chamber, which includes mode stirrers that are oriented to promote field homogeneity and has been studied primarily as a testing facility in the field of electromagnetic compatibility \cite{holloway2012reverberation,boyes2015reverberation}. The usage of ``RC" herein strictly refers to the resonant chamber, which is devoid of any mode stirrers and other test equipment. 

In this work, we assume that the boundary of the chamber is composed of a perfect electric conductor (PEC). Accounting for losses in the boundary is beyond the scope of this paper and will be explored in future work. However, we do provide a loss mechanism in the form of a load resistor that terminates the receive antenna. See Section \ref{losslessVSlossy} for further comments on loss in the system.

The RC has been studied extensively by the wave propagation community. Bethe and Schwinger pioneered cavity perturbation theory for optics \cite{bethe1943perturbation}, where they considered the effects of small changes to the cavity boundaries and introduction of small dielectric objects in the interior. Another point of interest has been in analyzing the field patterns inside the RC, via both a deterministic and statistical approach \cite{hill2009electromagnetic, collin1990field,
arnaut2006electromagnetic, morse1954methods}. The latter is necessitated by the extreme sensitivity to slight variations inside the RC. Naturally, this has produced papers that view the RC from a chaos theoretic perspective \cite{orjubin2007chaoticity, cappetta1998electromagnetic, legrand2010wave, lin2020stochastic}. 

With respect to antenna systems and communication specifically, a noise model was recently derived from classical statistical mechanics that takes the ambient noise in the RC into account \cite{10000780}. In \cite{boyes2015reverberation,xu2019anechoic,rosengren2005radiation}, it is mentioned how to compute Shannon capacity for antenna systems in an RC. However, the capacity expressions shown in these studies are meant to be solely of numerical and experimental use. In other words, the presentation of capacity in these works necessitates data collection.

The main contribution of this paper is to formulate a general theory for single-input, single-output (SISO) wireless communication where the channel is a lossless two-port network, the transmit port is driven by a current source, and the receive port is terminated in a load resistor. This model handles the RC, its 1D counterpart the transmission line, and the parallel inductor-capacitor network. The novelty of this model is two-fold: first, system poles can be very close to the real frequency axis (for an infinite load resistor, the poles are actually on the real frequency axis), and second, there is strong interaction between the transmitter and the receiver.

The paper develops as follows:
\begin{itemize}
    \item Section II contains detailed derivations of the physics of the parallel LC network and the transmission line segment, resulting in explicit expression for the $2 \times 2$ impedance matrix.
    \item Section III develops the Shannon theory for the lossless two-port network, where the transmit port is driven by a current source, and the receive port is terminated in a load resistor whose voltage is measured by an infinite-impedance amplifier. There are two sources of receiver noise: the Johnson noise of the load resistor, and the amplifier noise. Some surprising results emerge: first, for a fixed transmit power, capacity increases without bound as the load resistance increases; second, the capacity-attaining (``water-filling") allocation of power versus frequency avoids placing power close to the resonance frequencies.
    \item Section IV has detailed numerical results for the LC network and for the transmission line segment to illustrate the theory developed in the previous section. The qualitative features of the numerical results are quite similar for the two channels.
    \item Section V discusses the effects of going from a lossless two-port network to a lossy two-port network. We argue heuristically that the major features of loss are captured in the modeling of Section II by the appropriate choice of the load resistance.
    \item Section VI lists several follow-up research items to be pursued.
\end{itemize}



\section{Lossless Two-Port Networks}

A ported device, examples of which are resistors, capacitors, inductors, and antennas, is accessed electrically by a pair of wires that carry equal and opposite currents, and between which is a voltage. A collection of $N$ ported devices comprises an $N$-port network, characterized by a $N \times N$ complex-valued impedance matrix that relates the $N$ voltages to the $N$ currents \cite{ivrlavc2010toward,Youla}. The diagonal elements are the self-impedances. The off-diagonal elements are the mutual impedances, characterizing the interactions between pairs of devices. Reciprocity implies that the impedance matrix is non-conjugate symmetric. For a passive network, conservation of energy implies that the real part of the impedance matrix is non-negative definite.

A system comprising two antennas inside a lossless resonant chamber constitutes a two-port passive network having a $2 \times 2$ imaginary-valued impedance matrix. A transmission line is equivalent to a one-dimensional resonant chamber. A particularly simple example of a lossless two-port network is a parallel LC (inductor-capacitor) network.

\subsection{Parallel LC network}

\begin{figure}
\centering
\includegraphics{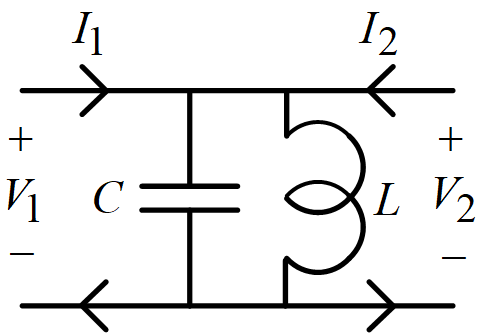}
\caption{Parallel LC two-port network: the system has two poles on the real frequency axis associated with the resonant frequency}
\label{LC2port}
\end{figure}

Fig. \ref{LC2port} shows a two-port network based on a parallel LC combination.

\begin{figure}
\centering
\includegraphics[width=3.49in]{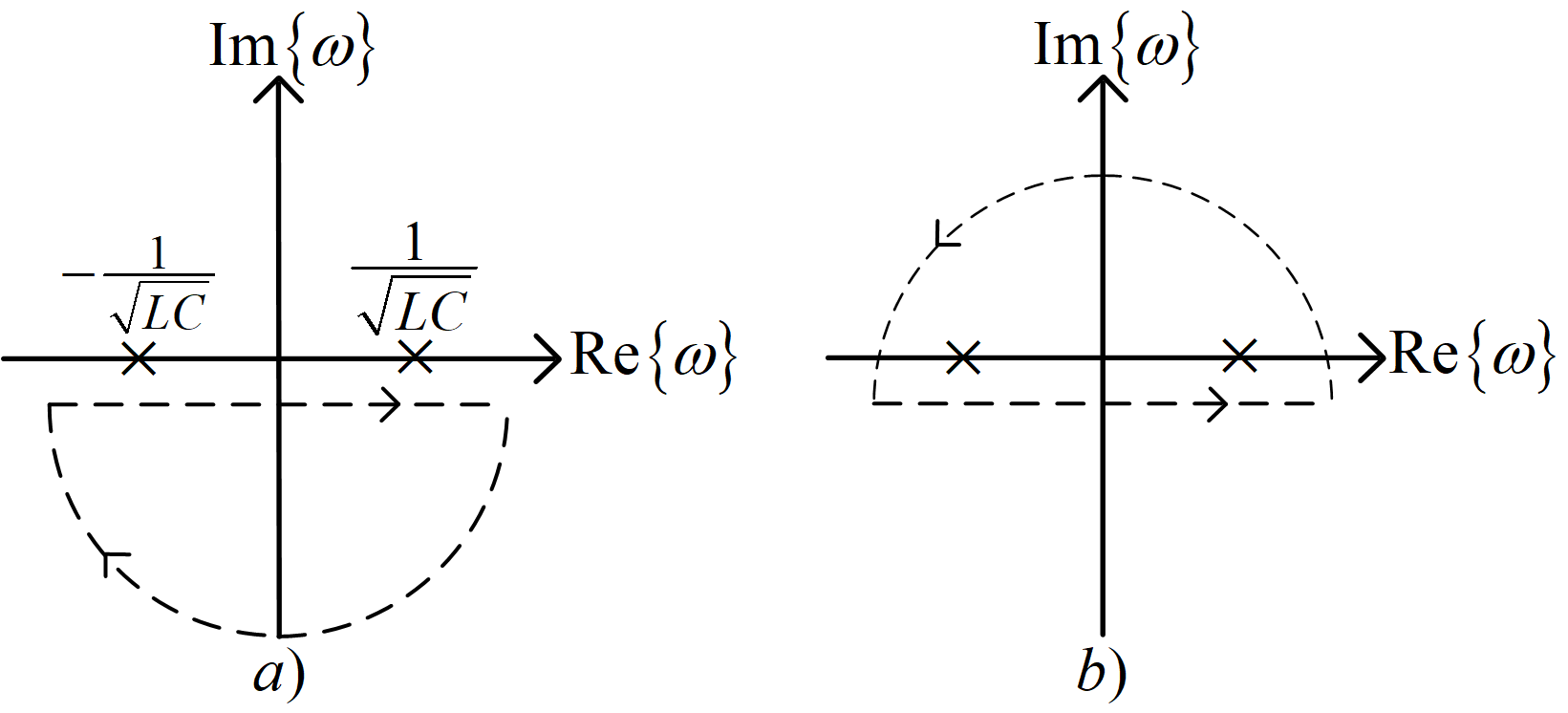}
\caption{Contours for evaluating inverse Fourier transform via calculus of residues: a) $t<0$, b) $t>0$}
\label{LCcontour}
\end{figure}

There is a simple voltage/current relation,
\begin{eqnarray} \label{RC}
V_1 = V_2 &=& \frac{I_1 + I_2}{\mathrm{i} \omega C + \frac{1}{\mathrm{i} \omega L}} \nonumber \\
&=& \frac{\mathrm{i} \omega L  \left( I_1 + I_2 \right)}{1 -  L C \omega^2} .
\end{eqnarray}
This translates into an impedance matrix which is symmetric and imaginary-valued,
\begin{equation} \label{LCZ}
\left[ \begin{array}{c} V_1 \\ V_2 \end{array}  \right] =  \frac{\mathrm{i} \omega L }{1 -  L C \omega^2} \left[ \begin{array}{cc} 1 & 1 \\ 1 & 1 \end{array}  \right] \left[ \begin{array}{c} I_1 \\ I_2 \end{array}  \right] .
\end{equation}
Viewed as a communication channel, the transmitter would drive the network with current $I_1$, and the receiver would measure the open-circuit voltage $V_2$,
\begin{equation}
\frac{V_2(\omega)}{I_1(\omega)} = Z_{21}(\omega) = \frac{\mathrm{i} \omega L }{1 -  L C \omega^2} .
\end{equation}
Note there are two poles on the real frequency axis, $\omega = \pm \frac{1}{\sqrt{LC}}$. The system is causal, so the region of convergence of the Fourier transform is the lower-half $\omega$-plane, $\mathrm{Im} \left\{ \omega \right\} \equiv \omega^{\prime \prime} < 0$. The inverse Fourier transform yields the impulse response, and can be evaluated by deforming the integration contour as shown in Fig. \ref{LCcontour}, and applying the Cauchy residue theorem for $\omega^{\prime \prime} < 0$,
\begin{eqnarray}
z_{21}(t) &=& \int_{- \infty + \mathrm{i} \omega^{\prime \prime}}^{ \infty + \mathrm{i} \omega^{\prime \prime}} \frac{\mathrm{d} \omega}{2 \pi} \frac{-\mathrm{i} \omega}{C \left(\omega - \frac{1}{\sqrt{LC}} \right)\left(\omega + \frac{1}{\sqrt{LC}} \right) } \mathrm{e}^{\mathrm{i} \omega t} \nonumber \\
&=& \frac{\cos \left(  \frac{t}{\sqrt{LC}} \right)}{C} \mathrm{u}(t), 
\end{eqnarray}
where $\mathrm{u}(t)$ is the unit step function.
\subsection{Transmission Line}

A transmission line has a distributed capacitance, $\bar{C}$ (F/m), and a distributed inductance, $\bar{L}$ (H/m), which support a distributed voltage and current. We also assume a distributed current source, $j(t,x)$ (A/m) such that,
\begin{eqnarray} \label{1Dvi}
    \frac{\partial v(t,x)}{\partial x} &+& \bar{L} \frac{\partial i(t,x)}{\partial t} = 0, \nonumber \\
    \frac{\partial i(t,x)}{\partial x} &+& \bar{C} \frac{\partial v(t,x)}{\partial t} = j(t,x) .
\end{eqnarray}
In turn, (\ref{1Dvi}) implies that both the voltage and the current satisfy the wave equation driven by the distributed current source,
\begin{eqnarray} \label{1Dwequ}
\frac{\partial^2 v(t,x)}{\partial x^2}- \frac{1}{c_0^2} \frac{\partial^2 v(t,x)}{\partial t^2} &=& - \frac{z_0}{c_0} \frac{\partial j(t,x)}{\partial t}, \nonumber \\
\frac{\partial^2 i(t,x)}{\partial x^2}- \frac{1}{c_0^2} \frac{\partial^2 i(t,x)}{\partial t^2} &=& \frac{\partial j(t,x)}{\partial x} ,
\end{eqnarray}
where $c_0$ is the propagation speed,
\begin{equation} \label{1Dc}
c_0 = \frac{1}{\sqrt{\bar{L} \bar{C}}} .
\end{equation}
The most general solution to the homogeneous wave equation, where the source-current is set to zero, is a superposition of plane-waves,
\begin{eqnarray} \label{1DPW}
v(t,x) &=& v_+ \left( t - \frac{x}{c_0}  \right) + v_- \left( t + \frac{x}{c_0} \right) , \nonumber \\
i(t,x) &=& y_0 v_+ \left( t - \frac{x}{c_0}  \right) - y_0 v_- \left( t + \frac{x}{c_0} \right) , 
\end{eqnarray}
for arbitrary waveforms $v_+(t), v_-(t)$; $y_0$ denotes the characteristic admittance, equal to the reciprocal of the characteristic impedance, $z_0$,
\begin{equation} \label{1Dz}
z_0 = \frac{1}{y_0} = \sqrt{\frac{\bar{L}}{\bar{C}}} .
\end{equation}

\subsubsection{Finite Length Transmission Line}
\begin{figure}
\centering
\includegraphics[width=3.49in]{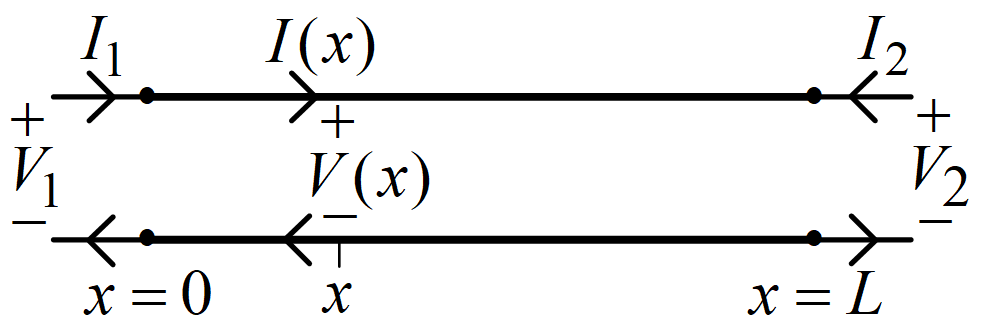}
\caption{Finite-length transmission line as a two-port network}
\label{TrLine_driven_ends}
\end{figure}

A finite-length transmission line, $0 \leq x \leq L$, constitutes a two-port network as shown in Fig. \ref{TrLine_driven_ends}.
In the following we derive the impedance matrix. The Fourier transform of (\ref{1DPW}) gives
\begin{equation} \label{VIvsV+-}
\left[ \begin{array}{c} V(\omega,x) \\ I(\omega,x) \end{array}  \right] = \left[  \begin{array}{cc} \mathrm{e}^{-\mathrm{i} k x} & \mathrm{e}^{+\mathrm{i} k x} \\ y_0 \mathrm{e}^{-\mathrm{i} k x} & - y_0\mathrm{e}^{+\mathrm{i} k x} \end{array} \right] \left[ \begin{array}{c} V_+(\omega ) \\ V_-(\omega) \end{array}  \right] ,
\end{equation}
\begin{equation} \label{k}
k = \frac{\omega}{c_0} = \frac{2 \pi}{\lambda} .
\end{equation}
With reference to Fig. 3, and (\ref{VIvsV+-}),
\begin{eqnarray} \label{1Dconstraints}
V_1(\omega) &=& V(\omega,0) = V_+(\omega) + V_-(\omega) \nonumber \\
V_2(\omega) &=& V(\omega,L) = V_+(\omega) \mathrm{e}^{-\mathrm{i}kL} + V_-(\omega) \mathrm{e}^{+\mathrm{i}kL} \nonumber \\
I_1(\omega) &=& I(\omega,0) = y_0 V_+(\omega) - y_0 V_-(\omega) \nonumber \\
I_2(\omega) \nonumber &=& - I(\omega,L) \\ &=& -y_0 V_+(\omega) \mathrm{e}^{-\mathrm{i}kL} + y_0 V_-(\omega) \mathrm{e}^{+\mathrm{i}kL} .
\end{eqnarray}
We solve the third and fourth equations to obtain the plane-wave amplitudes in terms of the port currents,
\begin{equation} \label{1DV+V-}
\left[ \begin{array}{c} V_+(\omega) \\ V_-(\omega) \end{array}  \right] = \frac{-\mathrm{i} z_0}{2 \sin kL} \left[  \begin{array}{cc} \mathrm{e}^{+\mathrm{i} k L} & 1 \\  \mathrm{e}^{-\mathrm{i} k L} & 1 \end{array} \right] \left[ \begin{array}{c} I_1(\omega) \\ I_2(\omega) \end{array}  \right] .
\end{equation}
We then substitute (\ref{1DV+V-}) into the first two equations of (\ref{1Dconstraints}) to obtain the port voltages in terms of the port currents, e.g. the $2 \times 2$ impedance matrix for the transmission line segment,
\begin{equation} \label{1DZ}
\left[ \begin{array}{c} V_1(\omega) \\ V_2(\omega) \end{array}  \right] = \frac{-\mathrm{i} z_0}{\sin kL} \left[  \begin{array}{cc} \cos kL & 1 \\ 1 & \cos kL \end{array} \right] \left[ \begin{array}{c} I_1(\omega ) \\ I_2(\omega) \end{array}  \right] .
\end{equation}
As expected, the impedance matrix is symmetric and imaginary-valued. Note that the magnitude of the mutual impedance is comparable to that of the self-impedance --- characteristic of tight coupling between the ports.

The impedance matrix (\ref{1DZ}) has an infinite number of singularities with respect to frequency,
\begin{equation} \label{1Dpoles}
    \sin kL = \ \sin \frac{\omega L}{c_0} = 0 \leftrightarrow \omega = \frac{\pi c_0 \ell}{L}, \ \ell = 0, \pm 1, \pm 2, \cdots .
\end{equation}
The singularities are associated with normal modes in the transmission line. To understand this better we drive port-1 with a current, $I_1(\omega)$, and set the other port-current to zero, $I_2(\omega) = 0$. The substitution of (\ref{1DV+V-})
into (\ref{VIvsV+-}) gives the transfer function between the port-current and the distributed voltage and current,
\begin{equation} \label{1DVIvsportI}
\left[ \begin{array}{c} \frac{V(\omega,x)}{I_1(\omega)} \\ \frac{I(\omega,x)}{I_1(\omega)} \end{array}  \right] = \frac{-\mathrm{i}}{\sin \left( \frac{\omega L}{c_0}  \right)} \left[ \begin{array}{c} z_0 \cos \left( \frac{\omega (L-x)}{c_0}  \right) \\ \mathrm{i} \sin \left( \frac{\omega (L-x)}{c_0}  \right) \end{array}  \right] .
\end{equation}
Again, the Cauchy residue theorem is the simplest way to take the inverse Fourier transforms, but the key step is the recognition that the singularities are actually simple poles. To see this, we expand the sine function about one of its zeros,
\begin{eqnarray} \label{1Dsp}
\sin \left( \frac{\omega L}{c_0}  \right) &=& \sin \left( \frac{L \left( \omega - \frac{\pi c_0 \ell}{L}  \right)}{c_0} + \pi \ell  \right) \nonumber \\
&=& (-1)^\ell \sin \left( \frac{L \left( \omega - \frac{\pi c_0 \ell}{L}  \right)}{c_0} \right) .
\end{eqnarray}
Using the definition of residue, the minus-one coefficient in the Laurent series expansion, we have
\begin{eqnarray} \label{1Dresidue}
\mathrm{Res} \left\{  \frac{1}{\sin \left( \frac{\omega L}{c_0}  \right)} \right\}_{\omega = \frac{\pi c_0 \ell}{L}} &=& \lim_{\omega \to \pi c_0 \ell/L} \frac{\omega - \frac{\pi c_0 \ell}{L}}{\sin \left( \frac{\omega L}{c_0} \right)} \nonumber \\
&=& \frac{(-1)^\ell c_0}{L} . 
\end{eqnarray}
The impulse responses associated with the transfer functions (\ref{1DVIvsportI}) are obtained by summing over the residues,
\begin{align} 
\begin{split}
&\left. \qquad \hspace{-3mm}\int_{\omega = -\infty + \mathrm{i}\omega^{\prime \prime}}^{\infty + \mathrm{i}\omega^{\prime \prime}} \frac{\mathrm{d}\omega}{2\pi} \left[\begin{array}{c} \frac{V(\omega,x)}{I_1(\omega)} \\ \frac{I(\omega,x)}{I_1(\omega)} \end{array}\right] \mathrm{e}^{\mathrm{i}\omega t} \right|_{\omega^{\prime \prime} < 0} \\
&= \left. \sum_{\ell = -\infty}^{\infty} \frac{(-1)^\ell c_0}{L} \left[\begin{array}{c} z_0 \cos\left(\frac{\omega (L-x)}{c_0}\right) \\ \mathrm{i} \sin\left(\frac{\omega (L-x)}{c_0}\right) \end{array}\right] \mathrm{e}^{\mathrm{i}\omega t} \mathrm{u}(t) \right|_{\omega = \frac{\pi c_0 \ell}{L}}
\end{split} \nonumber \\
&= \frac{c_0}{L} \left( \left[\begin{array}{c} z_0 \\ 0 \end{array}\right] + 2 \sum_{\ell = 1}^{\infty} \left[\begin{array}{c} z_0 \cos\left(\frac{\pi \ell x}{L}\right) \cos\left(\frac{\pi c_0 \ell t}{L}\right) \\ \sin\left(\frac{\pi \ell x}{L}\right) \sin\left(\frac{\pi c_0 \ell t}{L}\right) \end{array}\right] \right) \mathrm{u}(t).
\end{align}
The last expression represents the impulse responses in terms of normal (Sturm-Liouville) standing-wave modes. The application of the product-to-sum trigonometric identities for $\cos(a)\cos(b)$ and $\sin(a)\sin(b)$  yields an expression that represents the impulse responses in terms of plane-waves.

Still another solution technique takes place in the time-domain. The driving current at $x=0$ creates an impulsive plane-wave which, at time $t = L/c_0$, arrives at $x=L$, where it undergoes a reflection, travels back to $x=0$ for still another reflection, and so on. For the open-circuited line, the voltage reflection coefficient is plus-one, and the current reflection coefficient is minus-one. The complete space-time system response to the current source, $i_1(t) = \delta(t)$, is
\begin{align} \label{1Dtime-domain}
    v(t,x) &= z_0 \sum_{m=0}^{\infty} \Delta_t^+\left(x + 2Lm\right) + \Delta_t^-(x-2L(m+1))\\
    i(t,x) &= \sum_{m=0}^{\infty} \Delta_t^+(x + 2Lm) - \Delta_t^-(x-2L(m+1)),
\end{align}
where \begin{equation}
    \Delta_t^{\pm}(x) \coloneqq \delta\left(t \mp \frac{x}{c_0}\right).
\end{equation}
The equivalent temporal Fourier transforms are
\begin{eqnarray} \label{1Dtime-domainF}
V(\omega,x) &=& z_0 \sum_{m=0}^{\infty} \left[ \mathrm{e}^{-\mathrm{i} \omega \left( \frac{x+2Lm}{c_0}  \right)} + \mathrm{e}^{\mathrm{i} \omega \left( \frac{x - 2L(m+1)}{c_0}  \right)} \right] , \nonumber \\
I(\omega,x) &=& \sum_{m=0}^{\infty} \left[ \mathrm{e}^{-\mathrm{i} \omega \left( \frac{x+2Lm}{c_0}  \right)} - \mathrm{e}^{\mathrm{i} \omega \left( \frac{x - 2L(m+1)}{c_0}  \right)} \right] .
\end{eqnarray}
After summing the geometric series (which converge for $\mathrm{Im} \left\{ \omega \right\} < 0$) and simplifying, we once again obtain (\ref{1DVIvsportI}).

\begin{figure}
\centering
\includegraphics[width=3.49in]{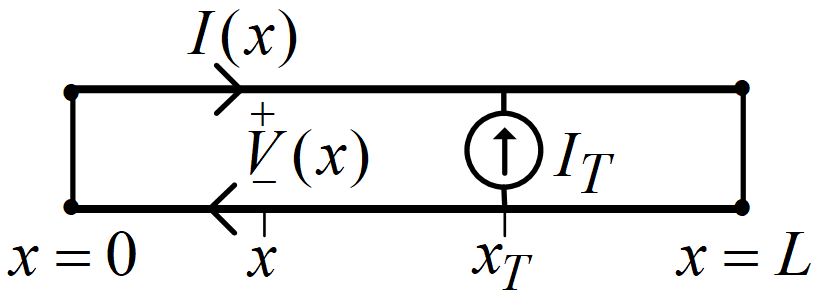}
\caption{Transmission line shorted at each end and driven by a current source at an intermediate point}
\label{TrLine_drivenXT}
\end{figure}

Fig. \ref{TrLine_drivenXT} represents a finite-length transmission line, short-circuited at both ends, and tapped at $x=x_T$ for a transmitter, and $x = x_R$ for a receiver.
We want to obtain the $2 \times 2$ impedance matrix that relates the transmit/receive voltages and currents. This is most simply done in the time-domain. The short-circuit voltage reflection coefficient is equal to minus-one, and the current reflection coefficient is equal to plus-one.

The current impulse initially creates left- and right-propagating plane-waves, 
\begin{equation}
    v(t,x) = \frac{z_0}{2} \delta\left(t - \frac{\bigl|x - x_{\mathrm{T}}\bigr|}{c_0} \right).
\end{equation}
The plane-waves are reflected at the short-circuit terminations, and the reflections are repeated at time intervals of $2L/c_0$. The complete time-domain solution is
\begin{align} \label{1Dv}
    v(t,x) &= \frac{z_0}{2}\Bigl\{ \Delta_t^+(|x - x_{\mathrm{T}}|) + \left[\Delta_t^+(x - x_{\mathrm{T}} + 2L) \right. \nonumber\\
    &+ \Delta_t^-(x - x_{\mathrm{T}} - 2L) -\Delta_t^+(x+x_{\mathrm{T}}) \nonumber\\
    &\left. - \Delta_t^-(x + x_{\mathrm{T}} - 2L) \right] * \sum_{m=0}^{\infty}\Delta_t^+(2Lm)\Bigr\},
\end{align}
where the asterisk denotes convolution.
After taking the Fourier transform, and summing the geometric series, we have 
\begin{equation} \label{1DxTv}
    V(\omega,x) = \mathrm{i} z_0\frac{\cos k \left( L - x - x_{\mathrm{T}} \right) - \cos k \bigl( L- \bigl| x - x_{\mathrm{T}} \bigr|  \bigr)}{2 \sin kL}.
\end{equation}
As a check, we note that $\left. V(\omega,x) \right| _{x=0,L} = 0$, and $\left. V(\omega,x) \right| _{x_{\mathrm{T}}=0,L} = 0$. An additional check is to verify directly that (\ref{1DxTv}) is the solution to the Helmholtz equation, i.e., the Fourier transform of (\ref{1Dwequ}).
The entries of the  $2 \times 2$ impedance matrix for the transmit and receive ports are
\begin{align} \label{1DshortZ}
Z_{\mathrm{T}} ( \omega) &= \frac{\mathrm{i} z_0}{2 \sin kL} \left[\cos k \left( L - 2 x_{\mathrm{T}}  \right) - \cos kL\right] \nonumber \\
Z_{\mathrm{R}} ( \omega) &= \frac{\mathrm{i} z_0}{2 \sin kL} \left[\cos k \left( L - 2 x_{\mathrm{R}}  \right) - \cos kL\right] \nonumber \\
Z_{\mathrm{RT}} ( \omega) &= \frac{\mathrm{i} z_0}{2 \sin kL} \left[\cos k \left( L - x_{\mathrm{R}} - x_{\mathrm{T}}  \right) \right. \nonumber \\
& \qquad \left. - \cos k \left( L- \left| x_{\mathrm{T}} - x_{\mathrm{R}} \right|  \right)\right] .
\end{align}


\subsection{3D Resonant Chamber}
We place a transmit antenna and a receive antenna inside a perfectly-conducting box, $0 \leq x \leq L$, $0 \leq y \leq L$, $0 \leq z \leq L$. A current in either of the antennas creates an electromagnetic field. The tangential component of the electric field has to vanish over the surfaces of the six walls. Techniques for obtaining a solution for the electromagnetic field in the box are based on the Sturm-Liouville expansion, and the plane-wave expansion \cite{morse1954methods}. The quantitative behavior of the associated $2 \times 2$ impedance matrix has features similar to that of the transmission line:
\begin{itemize}
\item The impedance matrix is pure-imaginary;
\item The magnitude of the mutual impedance and the self-impedances are comparable;
\item The box supports a countably infinite number of standing-wave normal modes, associated with single poles on the real frequency axis of the impedance matrix;
\item The impedance matrix changes significantly, as a function of space, on a scale of a half wave-length.
\end{itemize}

\subsection{Discussion}

The LC circuit, the finite transmission line, and the 3D resonant chamber are each described by a $2 \times 2$ impedance matrix, obtained from basic physics. For a lossless system, the impedance matrix is imaginary-valued, and it has simple poles on the real frequency axis. This generic model is the basis for the communication theory that we develop in the next section.

\section{Communication Through Lossless Two-Port Networks: General Theory}

In this section we develop a general theory for communication through a lossless two-port network. The channel between transmitter and receiver has unusual features, unprecedented in the communication theory literature: in particular the violent behavior with frequency because of poles on or near the real frequency axis, and the strong coupling between the transmit and receive antennas.

\subsection{System Description}

\begin{figure}
\centering
\includegraphics[width=3.49in]{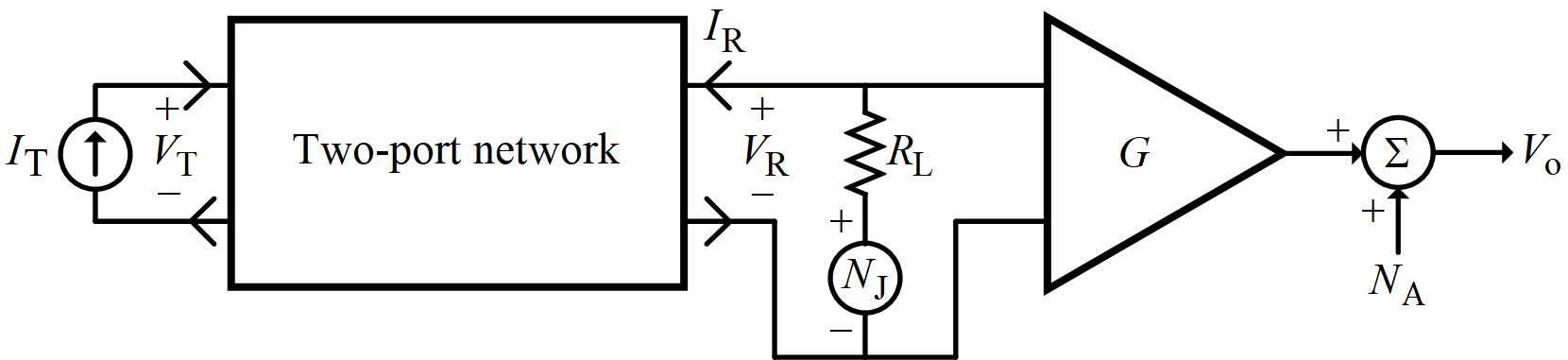}
\caption{Generic communication model for a lossless two-port network: a current source drives the transmit port, and the receive port is connected to a load resistor whose voltage is measured by an infinite-impedance amplifier; there are two sources of receiver noise: Johnson noise associated with the resistor, and the amplifier noise}
\label{Generic2port}
\end{figure}

Fig. \ref{Generic2port} illustrates a generic communication link for a lossless two-port network, described by an imaginary-valued impedance matrix,
\begin{equation} \label{V=ZI}
    \left[ \begin{array}{c}
    V_{\mathrm{T}}(\omega) \\
    V_{\mathrm{R}}(\omega)
    \end{array} \right] = 
    \left[ \begin{array}{cc}
    Z_{\mathrm{T}}(\omega) & Z_{\mathrm{TR}}(\omega) \\
    Z_{\mathrm{RT}}(\omega) & Z_{\mathrm{R}}(\omega) 
    \end{array} \right] 
    \left[ \begin{array}{c}
    I_{\mathrm{T}}(\omega) \\
    I_{\mathrm{R}}(\omega)
    \end{array} \right] .
\end{equation}
A current source, $I_{\mathrm{T}} (\omega)$, drives the transmit port. The receive port is connected to a load resistor, $R_{\mathrm{L}}$, and an infinite input impedance amplifier having gain $G$, and equivalent output additive noise, $N_{\mathrm{A}} (\omega)$. There is a second source of noise in the receiver: Johnson noise \cite{Bennett} which is associated with the load resistor, modeled as a series voltage source, $N_{\mathrm{J}} (\omega)$. Both the amplifier noise and the Johnson noise are modeled as white Gaussian,
\begin{eqnarray} \label{noise_ac}
\mathrm{E} \left\{ n_{\mathrm{A}} (t + \tau) n_{\mathrm{A}} (t)  \right\} &=& Q_{\mathrm{A}}\delta( \tau ), \nonumber \\
\mathrm{E} \left\{ n_{\mathrm{J}} (t + \tau) n_{\mathrm{J}} (t)  \right\} &=& 2 k_{\mathrm{B}} T R_{\mathrm{L}} \delta( \tau ),
\end{eqnarray}
where $k_{\mathrm{B}} = 1.38 \times 10^{-23}$ (J/K) is Boltzmann's constant, and $T$ (K) is absolute temperature; $Q_{\mathrm{A}}$ has units of $V^2$/Hz.

The transmitter operates subject to a bandwidth constraint of $B$ Hz, centered on a carrier frequency, $\omega_{\mathrm{c}}$,
\begin{equation}
    \omega_{c} - \pi B \leq | \omega | \leq \omega_{c} + \pi B ,
\end{equation}
and an expected power constraint,
\begin{equation} \label{power_constraint}
\mathrm{E} \left\{ i_{\mathrm{T}}(t) v_{\mathrm{T}}(t) \right\} \leq P_{\mathrm{T}} .
\end{equation}
Subject to these constraints, we want to find the capacity of the link, and the structure of the capacity attaining transmit current.

\subsection{Circuit Theory}

The load resistor imposes a constraint between the receive voltage and current (in the following, we suppress frequency dependence for the sake of brevity),
\begin{equation} \label{loadvi}
I_{\mathrm{R}} = \frac{N_{\mathrm{J}} - V_{\mathrm{R}}}{R_{\mathrm{L}}} .
\end{equation}
The substitution of (\ref{loadvi}) into the second equation of (\ref{V=ZI}) yields an expression for the receive voltage in terms of the transmit current and the Johnson noise voltage,
\begin{equation} \label{VR}
    V_{\mathrm{R}} = \frac{R_{\mathrm{L}} Z_{\mathrm{RT}} I_{\mathrm{T}} + Z_{\mathrm{R}} N_{\mathrm{J}} }{Z_{\mathrm{R}} + R_{\mathrm{L}}} .
\end{equation}
The substitution of (\ref{VR}) into (\ref{loadvi}) expresses the receive current in terms of the transmit current,
\begin{equation} \label{IR}
    I_{\mathrm{R}} = \frac{N_{\mathrm{J}} - Z_{\mathrm{RT}} I_{\mathrm{T}}}{Z_{\mathrm{R}} + R_{\mathrm{L}}} .
\end{equation}
In turn, the substitution of (\ref{IR}) into the first of (\ref{V=ZI}) yields an expression for the transmit voltage,
\begin{equation} \label{VT}
V_{\mathrm{T}} = \left( Z_{\mathrm{T}}
- \frac{Z_{\mathrm{RT}}^2 }{Z_{\mathrm{R}} + R_{\mathrm{L}}}  \right) 
I_{\mathrm{T}} + \frac{Z_{\mathrm{RT}} N_{\mathrm{J}}}{Z_{\mathrm{R}} + R_{\mathrm{L}}} .
\end{equation}
In conventional wireless links, the activities at the receiver have no measurable effect on the transmit voltage, and $V_{\mathrm{T}} \approx Z_{\mathrm{T}} I_{\mathrm{T}}$ --- a striking difference with respect to the lossless channel.

Again with reference to Fig. 5, the noisy output of the amplifier is

\begin{align} \label{noisy_signal}
   V_{\mathrm{o}}(\omega) &= GV_{\mathrm{R}}(\omega) + N_{\mathrm{A}}(\omega) \nonumber \\
   &= \underbrace{G\left(\frac{R_{\mathrm{L}}}{Z_{\mathrm{R}}(\omega) + R_{\mathrm{L}}}\right)Z_{\mathrm{RT}}(\omega) I_{\mathrm{T}}(\omega)}_{\text \: \mathrm{signal}} \nonumber \\&+
   \underbrace{G\left(\frac{Z_{\mathrm{R}}(\omega)}{Z_{\mathrm{R}}(\omega) + R_{\mathrm{L}}}\right)N_{\mathrm{J}}(\omega)}_{\mathrm{Johnson~noise}}  ~+ \underbrace{N_{\mathrm{A}}(\omega)}_{\mathrm{receiver ~noise}}. 
\end{align}
This is an additive Gaussian noise waveform channel, to which conventional Shannon theory applies.
\subsection{Shannon Theory}
The channel constitutes a linear time invariant system, and the additive noises are stationary and Gaussian. In the transform domain, therefore, the channel (\ref{noisy_signal}) is equivalent to independent parallel channels. As a consequence, the capacity attaining transmit current is Gaussian and independent from one frequency to another. In accordance with the spectral representation, the transmit current is a stationary Gaussian random process. All that remains is to maximize mutual information with respect to its spectral density, $S_{\mathrm{I_{\mathrm{T}}}}(\omega)$, subject to bandwidth and power constraints.

\subsubsection{Expected Transmit Power}

The expected transmit power is equal to $\mathrm{E} \left\{ i_{\mathrm{T}}(t) v_{\mathrm{T}}(t)   \right\}$. The transmit voltage depends, via (\ref{VT}), on both the transmit current, and the Johnson noise voltage which is statistically independent of the transmit current. The transmit voltage is related to the transmit current through an equivalent filter,
\begin{equation}
   F(\omega) = \left( Z_{\mathrm{T}}
- \frac{Z_{\mathrm{RT}}^2 }{Z_{\mathrm{R}} + R_{\mathrm{L}}}  \right) .
\end{equation}

The expected power is
\begin{eqnarray} \label{PT}
\mathrm{E} \left\{ i_{\mathrm{T}}(t) v_{\mathrm{T}}(t)   \right\} &=& \int \frac{\dom}{2 \pi} F(\omega) S_{\mathrm{I}_{\mathrm{T}}}(\omega)  \nonumber \\
&=& 2 \int_{\omega_{\mathrm{c}} - \pi B}^{\omega_{\mathrm{c}} + \pi B} \frac{\dom}{2 \pi} \mathrm{Re} \left\{ F(\omega) \right\} S_{\mathrm{I}_{\mathrm{T}}}(\omega) \nonumber \\
&=& \int_{\omega_{\mathrm{c}} - \pi B}^{\omega_{\mathrm{c}} + \pi B} \frac{\dom}{2 \pi} \beta(\omega) S_{I_{T}}(\omega), \nonumber \\
\beta(\omega) &=& \frac{2 R_{\mathrm{L}} Z_{\mathrm{RT}}''^2 (\omega) } {Z_{\mathrm{R}}''^2 (\omega) + R_{\mathrm{L}}^2  } ,
\end{eqnarray}
where the double-prime denotes ``imaginary part."

\subsubsection{Mutual Information}

To compute mutual information, we need first to find the signal-to-noise ratio (SNR) as a function of frequency. The expression for the noisy signal, (\ref{noisy_signal}), combined with (\ref{noise_ac}), gives the spectral density for the noisy signal (\ref{SVo}), the first term of which is useful signal, and the second and third terms are noise,
\begin{equation} \label{SVo}
    S_{ V_{\mathrm{o}} }(\omega) = \frac{  G^2  Z_{\mathrm{RT}}''^2 R_{\mathrm{L}}^2 S_{\mathrm{I}_{\mathrm{T}}}(\omega)  } {Z_{\mathrm{R}}''^2 + R_{\mathrm{L}}^2 } + \frac{ 2 G^2 k_{\mathrm{B}} T Z_{\mathrm{R}}''^2 R_{\mathrm{L}} } {Z_{\mathrm{R}}''^2 + R_{\mathrm{L}}^2 } + Q_{\mathrm{A}} .
\end{equation}
Consequently the SNR is
\begin{eqnarray} \label{SNRomega}
\rho(\omega) &=& \alpha(\omega) S_{\mathrm{I}_{\mathrm{T}}}(\omega) \nonumber \\
\alpha(\omega) &=& 
\frac{ G^2  Z_{\mathrm{RT}}''^2(\omega) R_{\mathrm{L}}^2  } { 2 G^2 k_{\mathrm{B}} T Z_{\mathrm{R}}''^2(\omega) R_{\mathrm{L}} + Q_{\mathrm{A}} \left( Z_{\mathrm{R}}''^2(\omega) + R_{\mathrm{L}}^2\right) },
\end{eqnarray}
and the mutual information between the transmit current and the output voltage of the receiver is
\begin{equation} \label{mut_inf}
I \left\{  i_{\mathrm{T}}(t) ; v_{\mathrm{o}}(t) \right\} = \int_{\omega_{\mathrm{c}} - \pi B}^{\omega_{\mathrm{c}} + \pi B}
\frac{\mathrm{d} \omega}{2 \pi} \log_2 \left[ 1 + \rho(\omega)  \right] .
\end{equation}

\subsection{Maximization of Mutual Information: Capacity}

The maximization of mutual information (\ref{mut_inf}), subject to the power constraint (\ref{PT}), yields capacity (bits/s):
\begin{eqnarray} \label{cap_max}
C &=& \max_{S_{I_{\mathrm{T}}}(\omega) \geq 0} \left\{ \int_{\omega_{\mathrm{c}} - \pi B}^{\omega_{\mathrm{c}} + \pi B}
\frac{\mathrm{d} \omega}{2 \pi} \log_2 \left[ 1 + \alpha(\omega) S_{I_{\mathrm{T}}}(\omega) \right] \right\}, \nonumber \\
&& \mathrm{subject \ to:} \ \int_{\omega_{\mathrm{c}} - \pi B}^{\omega_{\mathrm{c}} + \pi B}
\frac{\mathrm{d} \omega}{2 \pi} \beta(\omega) S_{I_{\mathrm{T}}}(\omega) \leq P_{\mathrm{T}} .
\end{eqnarray}
The application of the Kuhn-Tucker conditions yields the familiar ``water-filling" solution in terms of a positive-valued Lagrange multiplier, $\mu$, which is equal to the derivative of capacity with respect to transmit power.  A specific value of $\mu$ defines the support of the $S_{I{_\mathrm{T}}}(\omega)$,
\begin{equation} \label{Omega}
\Omega = \left\{  \omega: S_{I{_\mathrm{T}}}(\omega) > 0 \right\} \cap \left\{   \omega:  \omega_{c} - \pi B \leq | \omega | \leq \omega_{c} + \pi B \right\} ,
\end{equation}
as well as an explicit formula for $S_{I_{\mathrm{T}}}(\omega)$,
\begin{equation} \label{SI_mu}
    S_{I{_\mathrm{T}}}(\omega) = \left\{ \begin{array}{l} \frac{1}{\mu \beta(\omega)} - \frac{1}{\alpha(\omega)} > 0 , \ \omega \in \Omega  \\ 
    0, \ \omega \notin \Omega . \end{array} \right. 
\end{equation}
At frequencies where the ratio $\frac{\alpha(\omega)}{\beta(\omega)}$ falls below the Lagrange multiplier, transmit power is set to zero. The substitution of (\ref{SI_mu}) into (\ref{mut_inf}) and (\ref{PT}) gives the capacity and the transmit power in terms of the Lagrange multiplier,
\begin{equation} \label{C_general}
C = \int_{\omega \in \Omega} \frac{\mathrm{d} \omega}{2 \pi} \log_2 \left[  \frac{\alpha(\omega)}{\mu \beta(\omega)} \right] ,
\end{equation}
\begin{equation} \label{Pgeneral}
P_{\mathrm{T}} =  \int_{\omega \in \Omega} \frac{\mathrm{d} \omega}{2 \pi} \left[  \frac{1}{ \mu} - \frac{\beta(\omega)}{ \alpha(\omega)} \right] .
\end{equation}
In practice, one can assume a range of values of the Lagrange multiplier to obtain a cross-plot of capacity versus power.

\subsection{Upper and Lower Bounds on Capacity}

It is instructive to obtain upper and lower bounds on capacity.

\subsubsection{Upper Bound}

If we ignore the effect of the load resistor's Johnson noise, equivalent to assuming an absolute system temperature of zero, then we obtain an upper bound on capacity. For $T=0$ we have
\begin{equation} \label{alphT0}
\alpha(\omega) = \frac{ G^2  Z_{\mathrm{RT}}''^2(\omega) R_{\mathrm{L}}^2  } {  Q_{\mathrm{A}} \left( Z_{\mathrm{R}}''^2(\omega) + R_{\mathrm{L}}^2\right) },
\end{equation}
and according to (\ref{SI_mu}) we obtain a closed-form solution for the capacity-attaining spectral density, 
\begin{eqnarray} 
S_{I{_\mathrm{T}}}(\omega) &=& \frac{1}{\mu \beta(\omega)} - \frac{1}{\alpha(\omega)} \nonumber \\
&=& \frac{Z_{\mathrm{R}}''^2(\omega) + R_{\mathrm{L}}^2}{Z_{\mathrm{RT}}''^2(\omega) R_{\mathrm{L}}} \left(  \frac{1}{2 \mu} -\frac{Q_{\mathrm{A}}}{G^2 R_{\mathrm{L}} } \right)  .
\end{eqnarray}
By imposing the power constraint, we find that
\begin{equation} \label{SIupper}
S_{I{_\mathrm{T}}}(\omega) = \frac{P_{\mathrm{T}} \left( Z_{\mathrm{R}}''^2(\omega) + R_{\mathrm{L}}^2 \right)}{2B Z_{\mathrm{RT}}''^2(\omega) R_{\mathrm{L}}} , \ \left| \omega - \omega_{\mathrm{c}} \right| \leq \frac{B}{2}
\end{equation}
which gives a remarkably simple upper bound on capacity,
\begin{equation} \label{Cu}
    C < B \log_2 \left[ 1 + \frac{P_{\mathrm{T}} G^2 R_{\mathrm{L}} }{2B Q_{\mathrm{A}}}  \right] . 
\end{equation}

\subsubsection{Lower Bound}

To find a lower bound on capacity we compute mutual information for the now-suboptimal current spectral density, (\ref{SIupper}),

\begin{align} \label{Cl}
C > \int_{\omega_{\mathrm{c}} - \pi B}^{\omega_{\mathrm{c}} + \pi B}
\frac{\mathrm{d} \omega}{2 \pi} \log_2 \left[  1 + \frac{P_{\mathrm{T}} G^2 R_{\mathrm{L}} }{2B Q_{\mathrm{A
}}} \cdot \psi(\omega) \right],\\
\psi(\omega) = \left(  1 + \frac{2 G^2 k_{\mathrm{B}} T Z_{\mathrm{R}}''^2(\omega) R_{\mathrm{L}} }{Q_{\mathrm{A}} \left( Z_{\mathrm{R}}''^2(\omega) + R_{\mathrm{L}}^2\right)} \right)^{-1}.
\end{align}

\subsubsection{Discussion}

For a constant transmit power, $P_{\mathrm{T}}$, both the lower capacity bound (\ref{Cl}) and the capacity upper bound (\ref{Cu}) grow without limit as the load resistance $R_{\mathrm{L}}$ increases. This is intuitively reasonable: power is pumped into the system, and the only mechanism of power dissipation is through the load resistor. Therefore the mean-square voltage across the resistor is proportional to the resistance, which for a sufficiently high resistance, will overcome any amplifier noise.

The increasing load resistance is accompanied by a proportional increase in the Johnson noise spectrum. Paradoxically, the effect of the Johnson noise on the output signal \emph{decreases} with increasing resistance. We explain this by inspecting the second term of (\ref{noisy_signal}): the resistor and the self-impedance of the antenna form a voltage divider such that the relatively small self-impedance pulls down the voltage across the noisy resistor. Meanwhile, the r.m.s. Johnson noise voltage only increases as the square-root of the resistance: when divided by the resistance, the r.m.s. Johnson noise voltage at the input of the amplifier is inversely proportional to the square-root of the resistance.

The capacity upper bound (\ref{Cu}) has interesting implications for the link as the load resistance increases. In particular, capacity is independent of the carrier frequency and the positions of the transmitter and receiver.

The optimum distribution of transmit power vs. frequency, as given by the water-filling algorithm, has counter-intuitive behavior. In particular, water-filling \emph{avoids} placing power in the vicinity of the system poles. Both $\alpha(\omega)$ and $\beta(\omega)$ (\ref{SNRomega}), (\ref{PT}) have peaks in the proximity of poles. Placing power near a pole incurs a linear penalty in power, while only increasing capacity logarithmically. We can see this directly:
\begin{eqnarray} \label{alpha/beta}
   \frac{\alpha(\omega)}{\beta(\omega)} &=& \frac{ G^2   \left( Z_{\mathrm{R}}''^2 (\omega) + R_{\mathrm{L}}^2 \right)  R_{\mathrm{L}}   } {2 \left[ 2 G^2 k_{\mathrm{B}} T Z_{\mathrm{R}}''^2(\omega) R_{\mathrm{L}} + Q_{\mathrm{A}} \left( Z_{\mathrm{R}}''^2(\omega) + R_{\mathrm{L}}^2\right) \right] } \nonumber \\
   &=& \frac{ \left( \frac{G^2 R_{\mathrm{L}}}{2} \right)}{ \left( 2 G^2 k_{\mathrm{B}} T R_{\mathrm{L}} + Q_{\mathrm{A}} \right) - \left( \frac{2 G^2 k_{\mathrm{B}} T R_{\mathrm{L}}^3 d^2(\omega) }{\gamma^2(\omega) + R_{\mathrm{L}}^2 d^2(\omega)} \right)} \nonumber \\
   Z_{\mathrm{R}}''^2(\omega) &\equiv& \frac{\gamma^2(\omega)}{d^2(\omega)}.
\end{eqnarray}
By inspection we see that a pole such that $d(\omega) = 0$ is always a local minimum of the ratio $\alpha(\omega)/\beta(\omega)$. Consequently the optimum water-filling assigns power preferentially to frequencies away from the poles.

\section{Numerical Results: Capacity of Lossless Two-Port Networks}

In this section numerical results further illustrate the communication theory developed in the previous section.

For the computations we adopt some specific values for system parameters:
\begin{itemize}

\item 3 GHz carrier frequency, $\omega_{\mathrm{c}} = 2 \pi \cdot 3 \times 10^9$;

\item 10 MHz bandwidth, $B = 10^7$;

\item System noise temperature, $T = 300 ~\mathrm{K}$;

\item 40 dB amplifier gain, $G = 100$.

\end{itemize}
The amplifier noise density is calculated from the system noise temperature, referenced to 50 ohms, with an additional 9 dB spectral density,
\begin{eqnarray}
    Q_{\mathrm{A}} &=& 2 k_{\mathrm{B}} T \cdot 50 \cdot 10^{0.9} \nonumber \\
    &=& 3.29 \times 10^{-18}~  \mathrm{V^2/Hz}
\end{eqnarray}
where $k_{\mathrm{B}} = 1.38 \times 10^{-23}$ J/K.

Numerical quantities of interest include the deterministic transfer function between the transmit current and the receive voltage as given by (\ref{noisy_signal}), excluding the noise sources,
\begin{equation} \label{transfer_function}
\frac{V_{\mathrm{R}}(\omega)}{I_{\mathrm{T}}(\omega)} = \frac{ R_{\mathrm{L}} Z_{\mathrm{RT}}(\omega)  }{Z_{\mathrm{R}}(\omega) + R_{\mathrm{L}}} ,
\end{equation}
the ratio $\alpha(\omega)/ \beta(\omega)$ (\ref{alpha/beta}) which is central to the water-filling algorithm,
the capacity (\ref{C_general}) and the capacity upper and lower bounds, (\ref{Cu}), (\ref{Cl}), and the optimum transmit power density, (\ref{SI_mu}).

\subsection{Parallel LC Network}

We assume values for the capacitance and inductance so that the resonant frequency is $\omega_0/(2 \pi) = 3$ GHz: $C = 6.0 \times 10^{-13}$ F, $L = 4.7 \times 10^{-9}$ H.

\begin{figure}
\centering
\includegraphics[width=\columnwidth,trim = 1.5in 3in 1.5in 3in, clip]{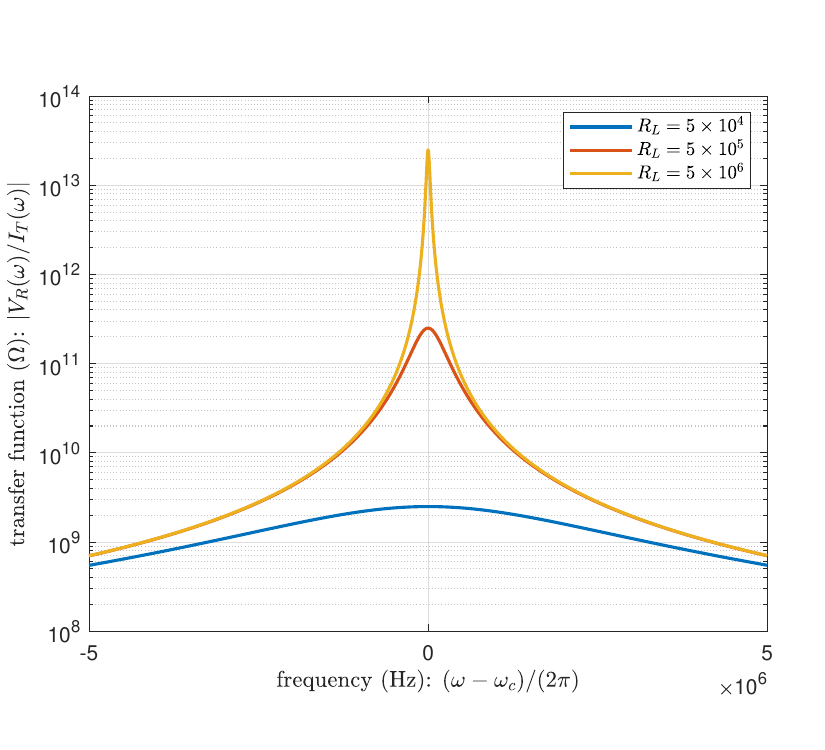}
\caption{LC circuit transfer function between transmit current and receive voltage for three values of the load resistor; a higher resistance moves the poles closer to the real frequency axis}
\label{LC_transfer}
\end{figure}
The transfer function between the transmit current and the receive voltage is plotted in Fig. \ref{LC_transfer} for three values of the load resistor, $R_{\mathrm{L}} = 5 \times 10^4, \, 5 \times 10^5, \ 5 \times 10^6$ ohms. A finite value of the load resistance moves the poles off the real axis. A greater value of the load resistance results in a sharper resonance.

\begin{figure}
\centering
\includegraphics[width=\columnwidth,trim = 1.5in 3in 1.5in 3in, clip]{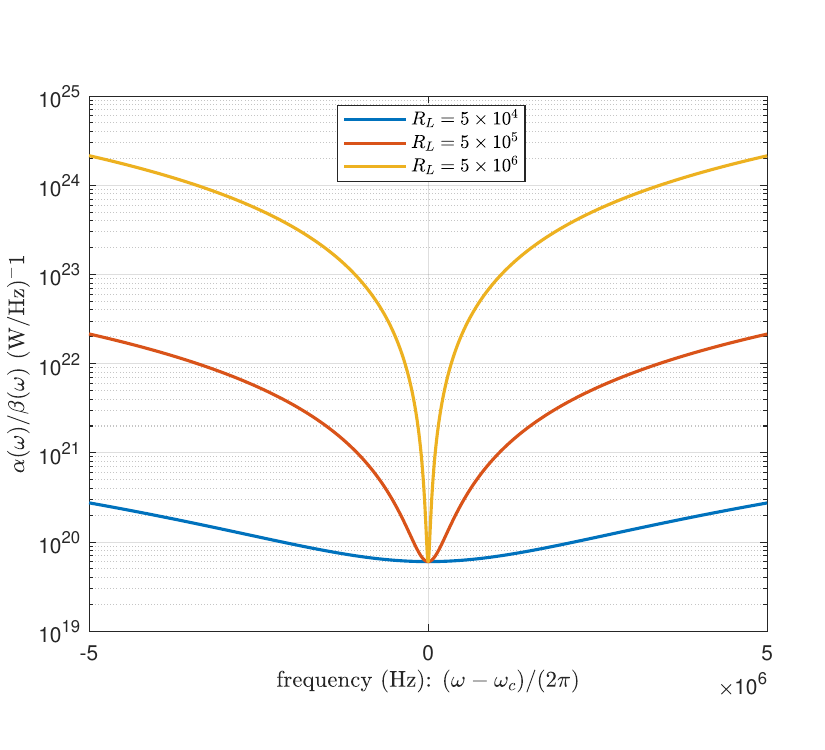}
\caption{LC circuit: the ratio $\alpha(\omega)/\beta(\omega)$; frequencies such that this ratio is greater than the Lagrange multiplier, $\mu$, are assigned transmit power --- as a result, frequencies near resonance (here $\omega = \omega_c$) are denied power}
\label{fig_7}
\end{figure}

The ratio $\alpha(\omega)/\beta(\omega)$ (\ref{alpha/beta}) is plotted in Fig. \ref{fig_7} for the three values of load resistance. Frequencies near the resonant frequency are denied power.

Fig. 8 displays a plot of spectral efficiency versus transmit power for the three values of the load resistor. The slope of the curve is proportional to the value of the Lagrange multiplier $\mu$ (steep for large $\mu$ and small power, shallow for small $\mu$ and large power). The curves are terminated at the point where $\alpha(\omega)/\beta(\omega) \geq \mu, \ \forall \omega \in \left[ \omega_{\mathrm{c}} - B/2, \omega_{\mathrm{c}} + B/2 \right]$, and all frequencies in the available spectrum are powered. Still greater power (associated with a smaller value of $\mu$) would be added uniformly over the frequency band via (\ref{Pgeneral}), but would yield only logarithmically increasing spectral efficiency.
\begin{figure}
\centering
\includegraphics[width=\columnwidth,trim = 1.5in 3in 1.5in 3in, clip]{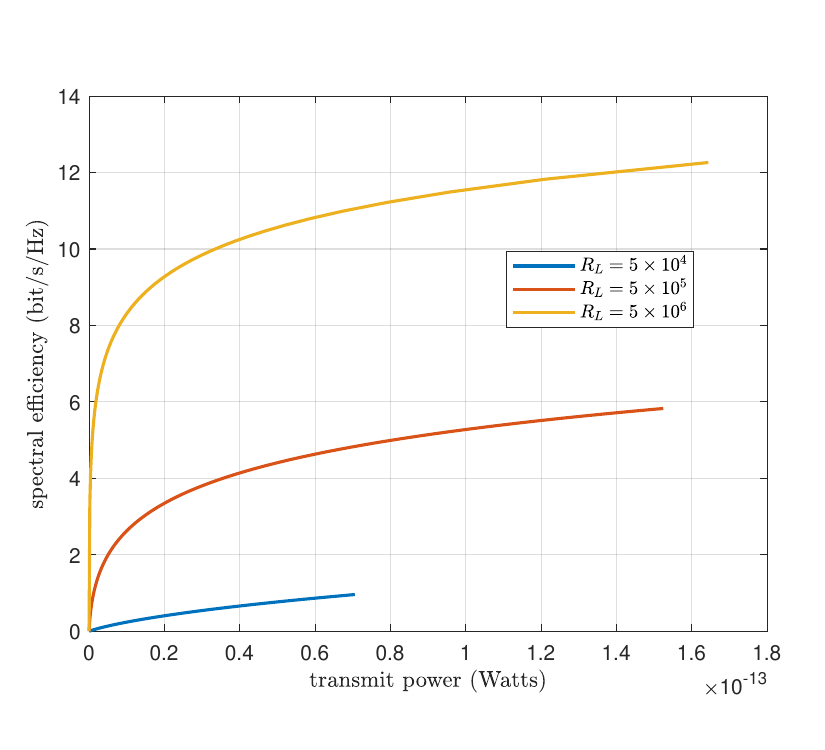}
\caption{LC circuit: spectral efficiency versus transmit power for three values of the load resistance. Each curve is terminated at the point where all frequencies in the available band are powered.}
\label{fig_8}
\end{figure}

Table 1 shows, for the three values of load resistance, the spectral efficiency, and its lower and upper bounds, for transmit power $P_{\mathrm{T}} = 2.68 \times 10^{-14}$ W (chosen to give spectral efficiency of 0.5 for the smallest value of load resistance). The lower bound is rather close to the true spectral efficiency. By ignoring the Johnson noise, the upper bound is overly optimistic.
\begin{table}[ht]
\caption{LC circuit: spectral efficiency, and its lower and upper bounds, for three load resistances, and for transmit power $P_{\mathrm{T}} = 2.68 \times 10^{-14}$ W}
\begin{center}
    \begin{tabular}{|c|c|c|c|}
    \hline
    $R_{\mathrm{L}} (\Omega)$ & lower bound & spectral efficiency (b/s/Hz) & upper bound \\
    \hline
    $5 \times 10^4$ & 0.426 & 0.500 & 17.6 \\
    \hline
    $5 \times 10^5$ & 3.58 & 3.67 & 21.0 \\
    \hline
    $5 \times 10^6$ & 9.69 & 9.70 & 24.3 \\
    \hline
    \end{tabular}
\end{center}
\end{table}

\begin{figure}
\centering
\includegraphics[width=\columnwidth,trim = 1.5in 3in 1.5in 3in, clip]{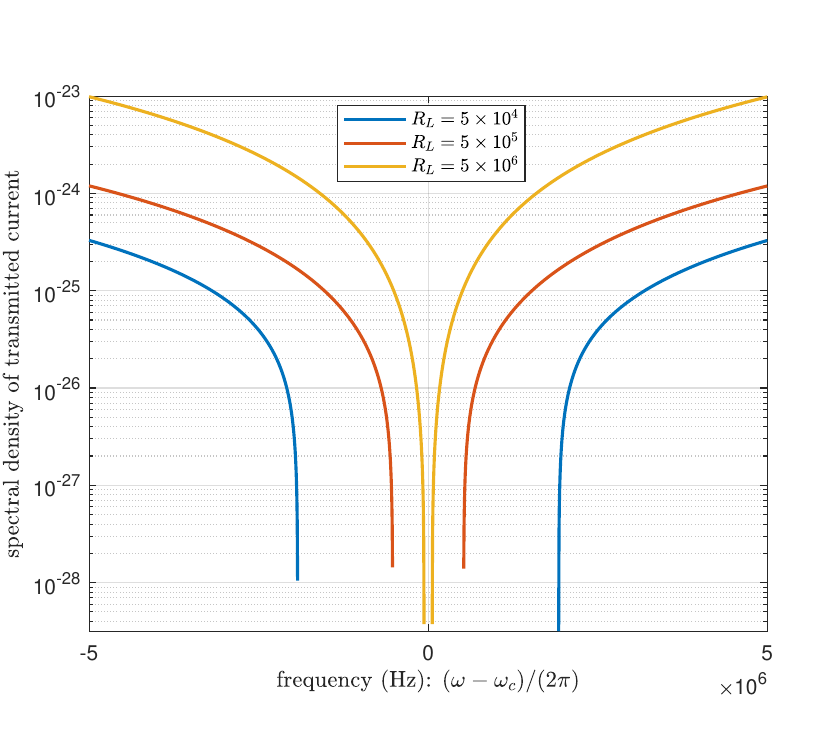}
\caption{LC circuit: capacity-attaining spectral density for transmit power $P_{\mathrm{T}} = 2.68 \times 10^{-14}$ W; frequencies near resonance are assigned no power}
\label{spectral_density_LC}
\end{figure}

Fig. \ref{spectral_density_LC} shows the capacity-attaining spectral density of the transmit current $S_{\mathrm{I}_{\mathrm{T}}}(\omega)$ for the three values of load resistance, corresponding to the transmit power $P_{\mathrm{T}} = 2.68 \times 10^{-14}$ W (e.g., Table 1).

\subsection{Transmission Line}
\begin{figure}
    \centering
    \includegraphics[width=\columnwidth,trim = 1.5in 3in 1.5in 3in, clip]{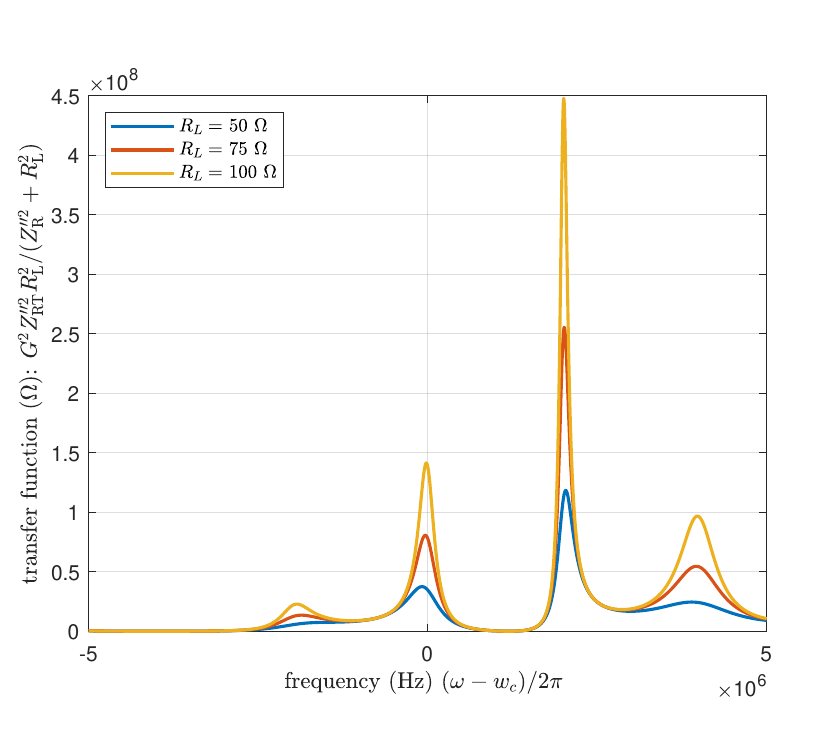}
    \caption{Transmission line transfer function between transmit current and receive voltage for three values of the load resistor; a higher resistance moves the poles closer to the real frequency axis}
    \label{fig:tlinetransferfunc}
\end{figure}

The transmission line, which is also a two-port lossless network, demonstrates very similar behavior to the LC network. 
For the following numerical results, we consider a transmission line that is $L = 75$ m in length. Using a carrier frequency of 3 GHz implies that this transmission line is approximately 750 wavelengths long. The transmitter and receiver positions $x_\mathrm{T}$ and $x_\mathrm{R}$ were set to $L/7$ m and $8L/13$ m, respectively.

Fig.~\ref{fig:tlinetransferfunc} shows the transfer function between the transmit current and receive voltage in the transmission line. One distinguishing feature of the transmission line compared to the LC network is that it has multiple resonances. Similar to the LC network, these resonances become sharper for greater values of the load resistance. 

\begin{figure}
    \centering
    \includegraphics[width=\columnwidth,trim = 1.5in 3in 1.5in 3in, clip]{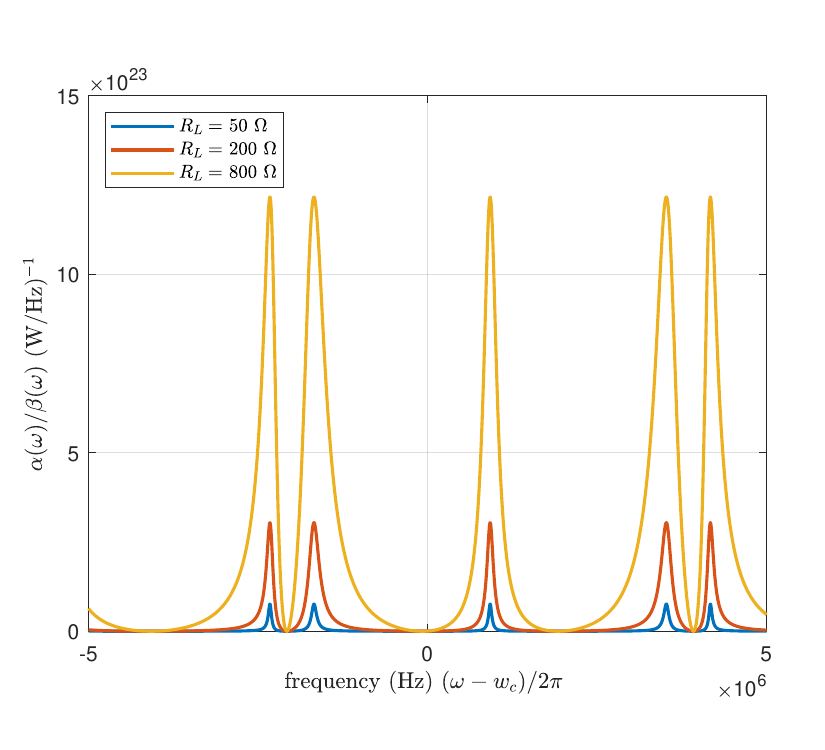}
    \caption{Transmission line: the ratio $\alpha(\omega)/\beta(\omega)$; frequencies such that this ratio is greater than the Lagrange multiplier, $\mu$, are assigned transmit power --- as a result, frequencies near resonance (here $\omega = \omega_c$) are denied power}
    \label{fig:alphabeta}
\end{figure}

\begin{figure}
    \centering
    \includegraphics[width=\columnwidth,trim = 1.5in 3in 1.5in 3in, clip]{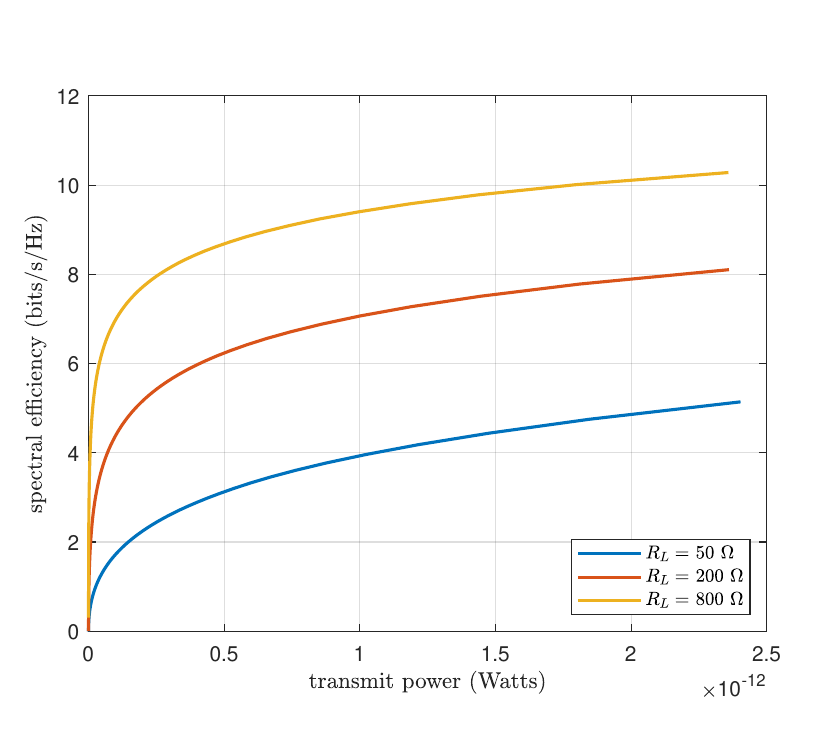}
    \caption{Transmission line: spectral efficiency versus transmit power for three values of the load resistance. Each curve is terminated at the point where all frequencies in the available band are powered.}
    \label{fig:tlinespeceff}
\end{figure}

\begin{figure}
    \centering
    \includegraphics[width=\columnwidth,trim = 1.5in 3in 1.5in 3in, clip]{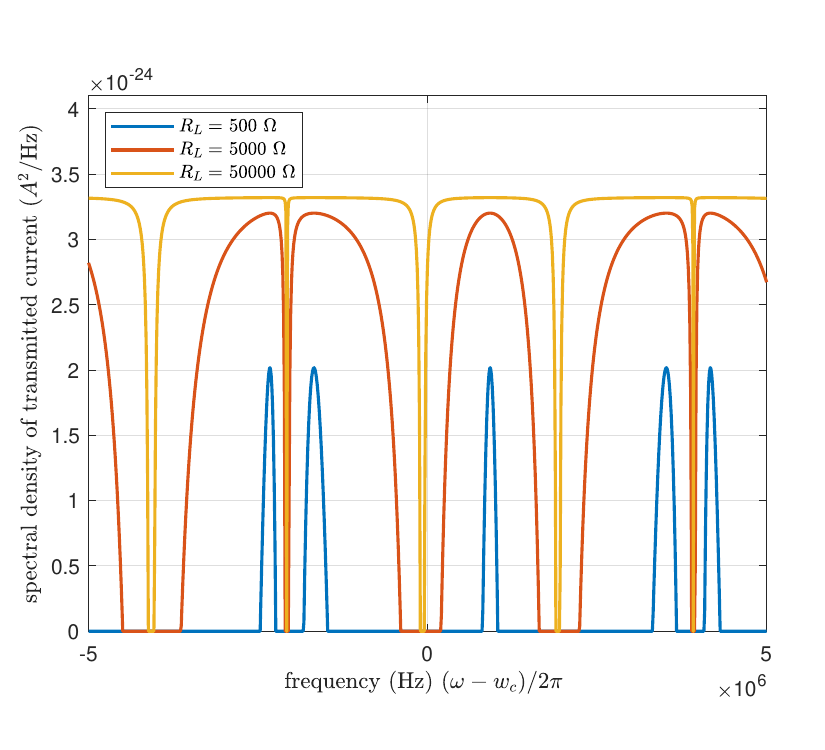}
    \caption{Transmission line: capacity-attaining spectral density; frequencies near resonance are assigned no power}
    \label{fig:tlinespecdens}
\end{figure}

Fig.~\ref{fig:alphabeta} shows the ratio $\alpha(\omega)/\beta(\omega)$ for the transmission line. Just as in the LC network case, we compare this ratio to the inverse of the Lagrange multiplier, i.e. $1/\mu$, at each frequency of interest to see how much power should be given there. An instance of spectral density, which is directly related to this ratio, is given for three different values of resistance in Fig.~\ref{fig:tlinespecdens}. As observed for the LC network, power is not distributed over the resonant frequencies and the spectral density $S_{I{_\mathrm{T}}}(\omega)$ drops sharply there.   Fig.~\ref{fig:tlinespeceff} illustrates spectral efficiency as a function of transmit power for three different load resistances.

\section{Lossless vs. Lossy Systems}
\label{losslessVSlossy}
The communication theory of section III and the numerical results of section IV assume a lossless two-port network where the receive antenna is terminated in a finite load resistor. It is reasonable to ask what happens when the system itself has loss.

In the case of a resonant chamber, loss would occur either if the walls of the chamber had non-zero resistivity, or if the interior of the chamber contained lossy objects, for example, pieces of metal having non-zero resistivity.

Loss would occur in the transmission line segment if the two terminations were finite resistors rather than short circuits. This corresponds to a RC with walls having non-zero resistivity. Alternatively there could be a distributed resistance within the transmission line itself.

Irrespective of the details of the loss mechanism of the two-port network, its effect would be qualitatively similar to that of terminating the receive antenna with a finite load resistance, which moves the system poles off the real frequency axis. In other words, an artful choice of the value of the load resistance could conceivably make the lossless network look like a lossy network. Certainly the precise manner in which loss occurs would affect the details of the frequency response of the system, but intuition suggests only minor difference in the ultimate performance of the communication link.
A further investigation is warranted, but is a topic for future research.

\section{Conclusions}

We have developed a theory for single-input, single-output (SISO) communications where the channel is described by a lossless two-port network. In our model, a loss mechanism is provided by terminating the receive antenna with a finite resistance. The two principal conclusions are first, for a constant transmit power, capacity is unbounded as the load resistance increases, and second, the capacity-attaining allocation of transmit power versus frequency typically avoids the assignment of power to frequencies near the system poles.
Our results suggest that a resonant chamber could potentially be an attractive artificial communication environment.

The topic of wireless communication in an RC offers many targets for future research, including:
\begin{itemize}
\item How can we realistically model the electromagnetic behavior of an RC? Analytical techniques can conceivably model a geometrically perfect box whose walls have non-zero resistivity, but anything else would require a purely numerical solution having extreme sensitivity to detailed assumptions. A statistical approach may be the only viable way to proceed.
\item Because of the long reverberation time of an RC, OFDM may not be viable because of the long cyclic prefix needed. A new modulation scheme is required that can handle very long delay-spread.
\item Multiple-antenna (MIMO) technology would certainly appear in practical RC communications, particularly because the RC constitutes a ``rich scattering" environment. The diversity provided by MIMO could be of great help should one or more antennas fall in a dead zone of the RC.
\item If one were to build a factory or a data center as an RC, how closely could it be made to approach the ideal lossless RC?
\end{itemize}

\section{Acknowledgements}
The authors would like to thank Sundeep Rangan at NYU Tandon School of Engineering and Thorkild B. Hansen at Seknion, Inc. for insightful discussions as well as the editors and reviewers for helpful feedback. 
\bibliographystyle{IEEEtran}
\bibliography{main.bib}
\end{document}